\documentclass[prl,superscriptaddress,aps,twocolumn,amsmath,nofootinbib]{revtex4}
\usepackage{epsfig}
\usepackage{graphics}
\usepackage{color}

\def\jpb{{J. Phys. B: At. Mol. Opt. Phys.}~}
\def\pra{{ Phys. Rev. A}~}

\def\prl{{ Phys. Rev. Lett.}~}
\def\jmo{{ J. Mod. Opt.}~}
\def\rmp{{ Rev. Mod. Phys.}~}

\def\jetp{{ J. Exp. Theor. Phys.}~}
\def\etal{{\em et al. }}

\newcommand{\vecr}{\mathbf{r}}

\newcommand{\veck}{\mathbf{k}}

\newcommand{\vece}{\mathbf{e}}

\newcommand{\vecE}{\mathbf{E}}

\newcommand{\beq}{\begin{equation}}
\newcommand{\eeq}{\end{equation}}

\begin{document}

\title{Focal-shape effects on the efficiency of the tunnel-ionization probe\\
for extreme laser intensities}

\author{M. F. Ciappina}
\affiliation{Institute of Physics of the ASCR, ELI-Beamlines project, Na Slovance 2, 182 21 Prague, Czech Republic}
\affiliation{ICFO - Institut de Ciencies Fotoniques, The Barcelona Institute of Science and Technology, Av. Carl Friedrich Gauss 3, 08860 Castelldefels (Barcelona), Spain}

\author{E. E. Peganov}
\affiliation{National Research Nuclear University MEPhI, Kashirskoe av.31, 115409 Moscow, Russia}

\author{S. V. Popruzhenko}
\affiliation{Prokhorov  General Physics Institute of the Russian Academy of Sciences, Vavilov Str. 38, 119991, Moscow, Russia}

\date{\today}


\begin{abstract}
We examine the effect of laser focusing on the potential of a recently discussed scheme [M.F. Ciappina {\em et al}, \pra {\bf 99}, 043405 (2019); Las. Phys. Lett. {\bf 17}, 025301 (2020)] for {\em in situ} determination of ultra-high intensities of electromagnetic radiation delivered by multi-petawatt laser facilities.
Using two model intensity distributions in the focus of a laser beam, we show how the resulting yields of highly charged ions generated in the process of multiple sequential tunneling of electrons from atoms, depend on the shape of these distributions.
Our findings lead to the conclusion that an accurate extraction of the peak laser intensity can be made either in the near-threshold regime when the production of the highest charge state happens only in a small part of the laser focus close to the point where the intensity is maximal, or through the determination of the points where the ion yields of close charges become equal.
We show that, for realistic parameters of the gas target, the number of ions generated in the central part of the focus in the threshold regime should be sufficient for a reliable measurement with highly sensitive time-of-flight detectors.
Although positions of the intersection points generally depend on the focal shape, they can be used to localize the peak intensity value in a certain interval. 
Additionally to this analysis, we discuss the method in comparison to other recently proposed approaches for direct measurement of extreme laser intensities.
\end{abstract}

\maketitle

\section{Introduction}

Interaction of high-power electromagnetic radiation with different forms of matter including free electrons, atoms, molecules, solids and plasmas has constantly remained in the scope of experimental and theoretical researches over more than 50 years, since the early days of laser physics. 
For fundamental phenomena which can be potentially observed in such interactions, the strength of electric and magnetic fields of a laser wave is the crucial parameter determining the interaction scenario and the very possibility to experimentally approach of various effects.
The continuous progress in the development and application of high-power laser sources provides a gradual growth in the maximal available field strength.
This already made possible to experimentally explore the nonlinear quantum dynamics of atomic, molecular and solid state systems in the field of intense laser radiation; the physics of relativistic and ultrarelativisc electron plasma and relativistic nonlinear optics, see \cite{ivanov-rmp09,agostini-adv12,karnakov-ufn15,bulanov-rmp09,dipiazza-rmp12,fedotov-cp15} and references therein for an overview of these research fields.
Currently, electromagnetic fields of peak strength $E_0\approx 10^{12}$V/cm can be achieved in infrared and optical femtosecond pulses generated by multi-terawatt (TW) and petawatt (PW) laser systems.
For linearly polarized radiation, this electric field corresponds to intensity 
\beq
{\cal I}=\frac{cE_0^2}{8\pi}\approx 10^{21}{\rm W/cm}^2~.
\label{I}
\eeq
Singular reports of even higher intensities $\simeq 10^{22}{\rm W/cm}^2$ have been published \cite{perry-ol99,bahk-ol04,yanovsky-oe08,sulf-22} although without receiving independent unambiguous confirmations.

Last years, several laser facilities of multi-PW power have been commissioned and should start operating in a near future.
The incomplete list of such laser sources includes Apollon in France \cite{apollon}, APRI \cite{APRI} and SULF \cite{SULF} in China, CAEP \cite{CAEP} in South Korea and ELI in Czech Republic, Hungary and Romania \cite{ELI,P3-ELI}.
These new facilities are expected to start experimental campaigns in a couple of years, forming a network of 10-PW class laser sources.
Projects of even more powerful, sub-exawatt lasers are currently under development \cite{XCELS}.
These new technical achievements will open the door to experiments with electromagnetic fields of intensity $10^{22}\div 10^{24}{\rm W/cm}^2$, allowing physicists entering the new so far unexplored regimes of laser-matter interaction, including radiation-dominated dynamics of isolated charges and plasmas, laser initiation of cascades of elementary particles and excitation of extremely strong magnetic fields, a plethora of laboratory astrophysics phenomena, etc. 
\cite{bulanov-rmp09,dipiazza-rmp12,fedotov-cp15}.

In view of these expectations, the problem of precise characterization of the electromagnetic field distribution in a laser focus, and in particular the determination of the intensity peak value, becomes of high practical importance.
At high powers of radiation, a direct measurement of the intensity distribution using cameras is apparently impossible, while extrapolation of results recorded in the low-power regime is questionable and can only be considered as an indirect, and generally uncontrollable, estimate.
Thus, it is commonly accepted that reliable methods for laser focus characterization in the high-power regime should employ observation of effects whose magnitude will be highly sensitive to the parameters of the laser pulse.

Last time, several proposals for experimental characterization of a laser focus at extreme intensities have been discussed.
The list of currently examined methods for {\em in situ} intensity measurement includes diagnostics based on multiple tunneling ionization of heavy atoms \cite{ciappina-pra19,ciappina-book19,ciappina-lpl20}, on scattering of laser light by electrons \cite{dipiazza-ol12,he-oe19,marklund-arxiv19}, and on ponderomotive scattering of electrons or protons in the laser focus \cite{mackenroth-njp19,marklund-arxiv19,vais-njp20}.
In this paper, we continue our analysis of the atomic diagnostics based on the observation of tunneling ionization of multi-electron atoms in a laser focus \cite{ciappina-pra19,ciappina-book19,ciappina-lpl20}.
The idea of the method is to employ the highly nonlinear dependence of the probability of tunneling ionization both on the amplitude of the applied electromagnetic field and on the ionization potential of an atomic level.
This nonlinearity appears in form of the celebrated tunneling exponent \cite{oppen-pr28} and allows to estimate the peak intensity by knowing the highest charge state produced in the laser focus filled by a low-density gas of high-$Z$ atoms, with $Z$ being the nucleus charge.
Noble gases commonly used in strong field experiments will be particularly convenient for such measurements.
For each charge state, there is a well defined value of the threshold intensity required to produce a noticeable amount of the respective ions.
In particular, observation of bare ions of ${\rm Ar}^{18+}$, ${\rm Kr}^{36+}$ and ${\rm Xe}^{54+}$ would allow proving that intensities $\approx 3\cdot 10^{21}{\rm W/cm}^2$, $3\cdot 10^{23}{\rm W/cm}^2$ and $2\cdot 10^{24}{\rm W/cm}^2$ have been correspondingly achieved \cite{ciappina-pra19,ciappina-lpl20}.
The idea of atomic diagnostics based on multiple ionization of noble gases was proposed and experimentally verified in \cite{walker-pra01,walker-josa03,yamakawa-pra03,yamakawa-jmo03,link-rsi06} for intensities in the range of $10^{16}\div 10^{19}{\rm W/cm}^2$.

In our earlier publications \cite{ciappina-pra19,ciappina-book19,ciappina-lpl20} the method of atomic diagnostics was theoretically examined with the emphasis on the measurement of ultrahigh intensities ${\cal I}>10^{21}$W/cm$^2$.
A brief summary of the obtained results can be found in the next section.
The aim of the present paper is to consider the effect of the focal distribution of laser intensity on the ionic signal which can be measured by a time-of-flight (TOF) detector of ions.
As was demonstrated in \cite{ciappina-lpl20}, the intensity dependence of the ionic signal results from an interplay between the tunneling exponent dependence on the laser field strength and the volume effect determined by the shape of the laser focus.
Here we present a more detailed consideration of this problem by exploring two focus shapes including an exact solution to the Maxwell equations which describes a tightly focused static laser beam \cite{narozhny-jetp00}.
Our results deliver two important messages: (a) if a TOF detector is sensitive enough as to register a few dozens of highly charged ions generated in the central part of the focus, and the gas density in the jet can be accurately controlled, this would allow determining the peak value of intensity with a reasonable accuracy independently on the particular form of the intensity distribution within the focal spot; (b) simultaneous observation of intersections between the intensity-dependent yield curves would allow extracting the peak value of intensity independently to the value obtained in (a) and on the value of the concentration in the atomic jet.

The paper is organized as follows.
In the next section we give a brief overview of the recently discussed methods for {\em in situ} measurement of the peak laser intensity and its distribution in the focal spot.
Subsection 2.3 presents a compact description of the atomic diagnostics based on the observation of tunneling ionization of multielectron atoms including results reported in our previous publications \cite{ciappina-pra19,ciappina-book19,ciappina-lpl20}.
In Section 3 we describe two models of the laser focus which will be later used to examine the focal-averaging effect.
Results of this examination are reported in Section 4.
The last section contains brief conclusions.
Atomic units are used throughout unless otherwise stated.

\section{Proposals for {\em in situ} measurement of laser intensity}

The task of measuring the peak value of laser intensity or its distribution in space imposes several significant requirements on the effects which can be potentially used.
In order to not generate an additional field comparable to that of the laser wave, the target has to be sufficiently rare, which implies gases at low pressure or electron or ion beams of low density.
The crucial requirement to any potential method for determination of the peak intensity value is that the observed effect should be sensitive to the local values of the electric and magnetic fields.
As an example, for an atom in a low-pressure gas environment, the probability of its ionization is determined by the electromagnetic fields in the very vicinity of the atom.
In contrast, momentum distributions of photoelectrons produced in ionization of this atom are formed not only by the dynamics of the ionization event, but also by the motion of the photoelectrons from the atom to the detector through the laser focus where they experience ponderomotive scattering.
Thus, for the realistic case of a large number of atoms distributed in the laser focus, unambiguous information on the intensity value can hardly be extracted from the photoelectron spectra.

The requirements of locality and low target density leave only few effects to consider as potential candidates for an {\em in situ} intensity probe.
So far, tunnel ionization \cite{walker-pra01,yamakawa-pra03,link-rsi06,ciappina-pra19}, nonlinear Thomson scattering \cite{dipiazza-ol12,he-oe19,marklund-arxiv19} and ponderomotive scattering of electrons \cite{mackenroth-njp19,marklund-arxiv19} and protons \cite{vais-njp20} were examined by virtue of reaching this objective.
Below we briefly describe these effects giving a more detailed account of the method based on tunnel ionization.

\subsection{Thomson scattering}

In the field of an intense electromagnetic wave the classical motion of a charged particle becomes nonlinear.
The nonlinearlity is governed by the relativistically invariant dimensionless intensity parameter
\beq
a_0=\frac{zE_0}{mc\omega}
\label{a0}
\eeq
with $z$ and $m$ being the particle charge number and mass, in atomic units $z=m=1$ for the electron. 
The onset of nonlinear and relativistic effects, $a_0=1$ corresponds to intensity ${\cal I}\approx 1.4\cdot 10^{18}$W/cm$^2$ for an electron in the field of linear polarization and $\lambda=1\mu{\rm m}$.
Thus for intensities of interest the electron motion is highly nonlinear and ultrarelativistic.
The former results in emission of harmonics of the laser frequency, the latter makes the instant angular distribution of emitted radiation highly peaked along the velocity vector.
Radiation of a charged particle in the field of an intense electromagnetic wave is known as nonlinear Thomson scattering (NTS), a classical limit $(\hbar\to 0)$ of nonlinear Compton scattering \cite{ritus-jetp64,dauIV}.
A detailed account of the theory of NTS can be found in and \cite{sarachik-prd70}.
Angular distributions and spectra of emitted radiation depend in particular on the laser field amplitude which make them of potential use for determining the peak intensity value.

In \cite{dipiazza-ol12} it was proposed to measure the intensity via the angular offset in the distribution of radiation emitted by an ultrarelativistic electron counter-propagating to a laser pulse under examination. 
It was shown that an intensity $\simeq 4\cdot 10^{22}$W/cm$^2$ can be inferred with an $\approx 10\%$ accuracy if an electron beam of energy $\varepsilon_0=23$MeV with a narrow $\simeq 5\%$ energy spread collides to the laser pulse, and the angular distribution offset is measured with accuracy $\simeq 1\%$.
The peak intensity can be inferred with higher credibility when both radiation of an electron beam and its scattering are simultaneously characterized.
This approach has been examined in \cite{marklund-arxiv19}.
It was shown there that a highly accurate measurement becomes possible when an electron beam with a gamma-factor $\gamma\gg a_0$ and a width $b\ll w_0$, where $w_0$ is the laser focal waist, collides to a counterpropagating laser pulse.
The simultaneous measurement of the radiation angular variance and of the mean electron energies before and after the interaction allows to extract the peak value of $a_0$ with accuracy $\approx 10\%$.
In the field amplitude range $5<a_0<150$, where the upper limit corresponds to an intensity $\approx 5\cdot 10^{22}$W/cm$^2$, this scheme has also been shown practically insensitive to the model chosen for the description of the radiation reaction effect.
The schemes of \cite{dipiazza-ol12} and \cite{marklund-arxiv19} potentially provide a high accuracy of the intensity measurement.
However, they employ an electron beam of controllable quality which may complicate experimental realizations.

In a monochromatic field, the frequency $\omega_s$ of the $s$-th harmonic emitted by an electron generally depends on the angle $\theta$ between the wave vectors of the photon and electromagnetic wave and on the intensity parameter (\ref{a0}) ~\cite{sarachik-prd70}:
\beq
\omega_{s}=\omega_{s}(a_0,\theta,\psi_0)~.
\label{thomson}
\eeq
Here the variable $\psi_0$ denotes the initial condition for electron's motion which essentially specifies the character of trajectory in the laboratory frame.
By knowing this initial condition and measuring the frequency $\omega_s$ at fixed emission angle $\theta$ one can extract the value of $a_0$ and, therefore, that of ${\cal I}=c^3\omega^2a_0^2/8\pi $.

This procedure was demonstrated in a recent experiment \cite{he-oe19}, where a peak value of intensity $\simeq 10^{18}$W/cm$^2$ was estimated from the measurement of the second harmonic shift. 
NTS of laser radiation proceeded on electrons generated in the focus due to strong-field ionization of molecular nitrogen at a backing pressure $\simeq 10^{-3}$mbar. 
Then, the Thomson radiation was separated from that resulting from recombination, and the second harmonic frequency profile in the direction perpendicular to that of the laser beam propagation was recorded.
The measured value of intensity appeared to be in fair agreement with estimates of the peak intensity extracted from images of the focal area, obtained at reduced laser power.

The main drawback of this method stems from the dependence of the frequency shift in Eq.~(\ref{thomson}) on the initial condition.
In \cite{he-oe19} where the electron plasma was created due to the tunneling ionization on the leading edge of the laser pulse the particular form (\ref{thomson}) was used corresponding to the electron initially at rest.
However, with increasing of the peak intensities and therefore of ionization potentials of ionic species, this assumption will become less relevant.
As a consequence, the relation (\ref{thomson}) becomes considerably dependent on the time instant, within the laser pulse duration, at which the electron has been released.
This will make the procedure of intensity extraction from the spectra less unambiguous than it was in \cite{he-oe19}.
Another restriction of this method results from the small focal spot size and short pulse duration required to reach extreme intensities of electromagnetic radiation: in a non-monochromatic field the broadening of NTS spectra can considerably hinder determination of the intensity-induced shift.
This would be exactly the case for intensities $10^{23}$W/cm$^2$ and higher, which can only be reached through tight focusing with a waist of a few microns.
Then an electron will move in a field whose spatial inhomogeneity scale is comparable to the laser wavelength.
Complications in the intensity extraction algorithm rising with intensity increasing are discussed in \cite{he-oe19}
Finally, at ultrahigh intensities $>10^{25}$W/cm$^2$, NTS from protons \cite{he-oe19} can be used for the same purpose.

\subsection{Scattering of electrons and protons on the laser beam}
The intensity distribution in a laser focus can also be probed through scattering of charged particles.
This approach has been theoretically examined for electrons and protons, see e.g. recent publications \cite{vais-apb18,mackenroth-njp19,vais-njp20} and references there.
Scattering of charged particles on a laser focus in the limit of low density, when the self field of the beam makes no effect on the particle dynamics and radiation, have been widely considered starting from the pioneering work by Kibble \cite{kibble-pr66}.
A detailed account of the theory of relativistic ponderomotive scattering can be found in \cite{gor-josab89,narozhny-jetp00}.
Generally, there are no doubts that momentum distributions of electrons or protons after their interaction with a laser beam carry a complete information on the peak intensity value and on the intensity distribution in space and time.
The challenge is to find a suitable setup where the inverse problem of extraction of these values from the electron spectra can be reliably solved.
Scattering of electron bunches on a laser focus can give an information on the peak intensity value through the measurement of the maximal deflection angle of the electrons \cite{mackenroth-njp19}.
This approach suffers a serious drawback: a certain spatial profile of the laser has to be assumed to calculate the angle.
Besides, additional constraints are imposed on the parameters of the electron beam introducing extra difficulties into the experimental scheme and increases uncertainties in the interpretation of results.
In \cite{vais-njp20} the laser pulse diagnostics, based on the detection of momentum distributions of protons produced in ionization of a rarified hydrogen target and ponderomotively accelerated from the laser focus, was theoretically examined.
In the nonrelativistic domain of intensities for protons, ${\cal I}\le 10^{24}$W/cm$^2$ the method was shown to reliably determine the peak value of intensity via the cut-off position in the energy spectra of protons.
Moreover, the focal spot size was shown to be connected to the angular width of the proton spectra.

\subsection{Tunnel ionization of heavy atoms}
Nonlinear ionization of atoms or ions and related phenomena have been used as a probe of laser intensity for already several decades.
At moderate intensities, ${\cal I}\approx 10^{13}\div 10^{15}$W/cm$^2$, the single-electron above-threshold ionization (ATI) and the generation of high-order harmonics (HHG) are routinely observed processes.
The theory of ATI and HHG has been advanced to the quantitatively accurate level (see \cite{bauer-jpb06,ivanov-rmp09,poprz-jpb14,paulus-jpb18} and references therein for reviews) which allows extraction of the peak laser intensity from the position of the cut-off in HHG and high-energy ATI spectra or other characteristic features of photoelectron momentum distributions (for the latter, see e.g. \cite{kuebel-jpb18}) with $\sim 10\%$ accuracy. 
Measurements of photoelectron spectra have also been shown to allow the full characterization of laser pulses including their carrier envelope phase \cite{strelkov-oe14}.
In determination of the peak intensity, accuracy up to $1\%$ can be achieved by comparing experimental photoelectron yields from atomic hydrogen with predictions from exact numerical solutions of the three-dimensional time-dependent Schr\"odinger equation \cite{pullen-pra13}.
At higher intensities, HHG and high-order ATI vanish, and photoelectron distributions become affected by ponderomotive scattering as was discussed above.
The degree of ionization of an atomic specie becomes then the simplest measure of the peak intensity.
A fundamental advantage of nonlinear ionization as a probe for laser intensity stems from the relative physical simplicity of the process: the higher the intensity is, the smaller are the effects of electron-electron correlations and nonadiabaticity, owing to the smallness of the Keldysh parameter \cite{keldysh}
\beq
\gamma=\frac{\sqrt{2I_p}\omega}{E_0}\sim{\cal I}^{-1/2}~,
\label{gamma}
\eeq
where $I_p$ is the ionization potential of the bound state.
For intensities ${\cal I}>10^{15}$W/cm$^{2}$ and wavelengths in the infrared range, $\lambda\simeq 1\mu{\rm m}$, ionization proceeds in the regime $\gamma<1$, known as optical tunneling.
In the intensity domain ${\cal I}>10^{20}$W/cm$^{2}$ we are interested in here, $\gamma\ll 1$ electromagnetic fields of the laser wave can be considered static during the event of tunneling which makes formulas for the probability of quasi-static tunneling (known as Perelomov-Popov-Terentiev -- PPT rates \cite{ppt-jetp66a,pp-jetp67}) quantitatively applicable for description of multiple sequential ionization of heavy atoms. 
The validity of these formulas at relativistic intensities have been proven by measurements of the intensity dependence of ion yields recorded in ionization of noble gases at ${\cal I}\simeq 10^{19}$W/cm$^2$ \cite{walker-pra01,walker-josa03,yamakawa-pra03,yamakawa-jmo03}.
These measurements aimed to check the accuracy of the PPT tunneling formulas, but at the same time they offered a way to precisely measure high laser intensities using multiple tunneling ionization of heavy atoms.
In a following publication \cite{link-rsi06} the method for an {\em in situ} peak intensity measurement was experimentally verified at the Sandia Z PW laser facility.
There the basic principle of the method was clearly formulated as ``Detection of highest ionization charge state of a species and nondetection of next highest charge state essentially impose a lower and an upper bound on peak intensity, when the measurement is performed with a highly efficient detection system''.
It was demonstrated that a sufficient number of multiply charged ions can be detected in the single-shot regime, which is of particular importance for laser sources of ultrahigh power operating at low repetition rates.

The scheme of \cite{walker-pra01,walker-josa03,yamakawa-pra03,yamakawa-jmo03,link-rsi06} was examined in \cite{ciappina-pra19,ciappina-book19,ciappina-lpl20} aiming to adapt it for the diagnostics of extremely intense laser pulses of the future multi-PW laser facilities.
It was shown that:
\begin{itemize}
\item{Tunneling proceeds in the nonrelativistic regime up to intensities ${\cal I}\simeq 10^{26}$W/cm$^2$, so that nonrelativistic PPT rates can be safely used for calculations \cite{ciappina-pra19}. 
This statement must not be confused to the fact that {\em after} the tunneling the electron motion in such intense fields quickly becomes ultra relativistic.
This relativistic stage affects only the photoelectron distributions and not the ionic yields.}
\item{Although for heavy atoms and high charge ionic states the system of rate equations describing ionization cascades, which develop from a neutral atom, appear extremely cumbersome, it was shown that owing to the highly nonlinear dependence of the tunneling ionization rates on the values of ionization potentials and on the laser field amplitude, this system can be greatly reduced up to that containing only about a dozen of equations \cite{ciappina-pra19}. 
This allows to considerably accelerate numerical calculations.}
\item{The exponential dependence of the tunneling rates on the values of ionization potential $I_p$ and field amplitude $E$ allows to derive a simple and yet reliable estimate for the threshold intensity at which a given level is quickly ionized.
This condition reads $E_{*}\simeq 0.05(2I_p)^{3/2}$ \cite{ciappina-pra19} and can be used to select a suitable target for probing a certain intensity interval.}
\item{There is a danger that ionic states chosen for intensity diagnostics will not be fully saturated in the tunneling regime of ionization, but, with a further increase of intensity, the ionization process will partially proceed in the barrier suppression regime.
Barrier suppression ionization (BSI) is not covered by such quantitatively precise theory as tunneling ionization (for an overview of approximate methods for description of BSI see \cite{popov-usp04,lin-jpb05,kostyukov-pra18} and references there), so that entering the BSI regime can considerably decrease the accuracy of the method.
In order to avoid this pitfall, ionization of H- and He-like ions should be preferably used, which imposes an additional constraint on the atomic constituents of the target \cite{ciappina-lpl20}.}
\item{Finally, the focal volume effect was briefly examined in \cite{ciappina-lpl20}.
It was shown that the focal-integrated ionic signal demonstrates two characteristic features which can be used for determination of the peak intensity value, and, to some extent, of the focal shape.
This includes (a) the position of a steep part of the intensity-dependent yield, where the number of ions per laser shot increases from $N\simeq 1$ to $N\simeq 10^2$ while the peak intensity grows by $30\div 50\%$ and (b) the intersection point of the numbers of two neighboring ionic states: $N(A^{n+})\approx N(A^{(n+1)+})$.}
\end{itemize}

In the following we present a further analysis of the focal volume effect on the accuracy of intensity measurement.

\section{Calculation of the focal-averaged ionic signal}
In the nonrelativistic regime of tunneling, which correctly describes production of ionic states at intensities up to $\sim 10^{26}$W/cm$^2$ \cite{ciappina-pra19}, ionization rates depend only on the absolute value of the electric field $E(t)$, while the magnetic field makes no effect.
In a focused laser pulse the electric and magnetic field vectors are no longer orthogonal and equal in absolute values.
This means that the ionization atomic diagnostics will measure not the absolute value of the Poynting vector, but the value of $E_0$.
In order to avoid confusions, below we refer to the value of (\ref{I}) as to {\em effective} intensity.

For a given time dependence of the electric field $\vecE(\vecr,t)$, the populations $c_k(\vecr,t)$ of ionic states $A^{k+}$ can be found by solving numerically the system of rate equations with time-dependent coefficients expressed as linear combination of tunneling ionization rates.
These systems for argon, krypton and xenon, properly truncated from the side of small $k$ can be found in \cite{ciappina-pra19}.
When the coefficients $c_k$ are computed, their values at the end of the laser pulse, $t=T$, give the distribution $C_k(\vecr)\equiv c_k(\vecr,T)$ over the charge states $A^{k+}$ in a fixed space point.
The number of ions of charge $0\le k\le Z$ produced in the focal volume is then given by the integral
\beq
N(A^{k+})=n_0\int C_k(\vecr)d^3r~
\label{N}
\eeq
with $n_0$ being the initial concentration of neutral atoms.
In the coefficients $C_k$, the $\vecr$-dependence enters via that of the electric field amplitude or, equivalently, effective intensity (\ref{I}), so that $C_k$ can be considered as functions of the latter.
These functions have been numerically calculated in \cite{ciappina-pra19} for the pulse time envelope $\sin^2(\pi t/T)$, with a full duration $T=33$ fs, corresponding to 10 periods at $\lambda=1\;\mu{\rm m}$.
The intensity intervals were chosen to be $10^{19}\div 10^{22}$W/cm$^2$ for ${\rm Ar}^{14+}\div {\rm Ar}^{18+}$ and ${\rm Kr}^{26+}\div {\rm Kr}^{34+}$ and $10^{21}\div 5\cdot 10^{24}$W/cm$^2$ for ${\rm Xe}^{50+}\div {\rm Xe}^{54+}$.
Ionization potentials were taken from \cite{saloman-Ar,saloman-Kr,saloman-Xe}.

In order to examine a potential influence of the focal distribution distribution on the ion yield including its dependence on the peak intensity value ${\cal I}_{\rm m}$ we consider two model pulse shapes, viz. the simplest ${\rm TEM}_{00}$ Gaussian beam and an exact solution to the Maxwell equations for a stationary focused beam found in ~\cite{narozhny-jetp00}.

\subsection{Gaussian beam}
For the ${\rm TEM}_{00}$ Gaussian beam the intensity distribution results:
\beq
{\cal I}(x,y,z)=\frac{{\cal I}_{\rm m}}{1+z^2/z_R^2}\exp\bigg(-\frac{2(x^2+y^2)}{w_0^2[1+z^2/z_R^2]}\bigg)~.
\label{I-gauss}
\eeq
Here $w_0$ is the beam waist and $z_R=\pi w_0^2/\lambda$ is the Rayleigh length.
For numerical calculations, we take $w_0=3\;\mu{\rm m}$, so that for $\lambda=1\;\mu{\rm m}$, $z_R\approx 28.3\mu{\rm m}$, the FWHM focal spot size $d_0=\sqrt{2\ln 2}\:w_0\approx 3.5\;\mu{\rm m}$ and the beam angular divergence $\theta={\rm arctan}(\lambda/\pi w_0)\approx 0.105$.
At these parameters the value of $d_0$ is close to that reported in \cite{sulf-22} for the 10-PW SULF laser facility with the focused ability improved by optimizing the wavefront aberrations. 
The ion yield $N(A^{k+})$ is most sensitive to the value of ${\cal I}_{\rm m}$ near the ionization threshold of the $(k-1)$-th charge state where ${\cal I}_{\rm m}\approx {\cal I}_*(A^{k+})$ \cite{ciappina-pra19,ciappina-lpl20}.
In this domain, the ions are predominantly generated in the close proximity of the field maximum, where the distribution can be approximated by the parabolic expansion in (\ref{I-gauss}).
In this limit, the volume with intensities ${\cal I}_0\le {\cal I}\le {\cal I}_{\rm m}$ grows as
\beq
V(\xi)\approx\frac{2\pi}{3}\frac{w_0^3}{\theta}\xi^3~,~~~\xi=1-\frac{{\cal I}_0}{{\cal I}_{\rm m}}~.
\label{V-gauss}
\eeq

\subsection{Narozhny-Fofanov beam}
We will compare results obtained for the beam (\ref{I-gauss}) with those for another focused beam, which is an exact stationary solution to the Maxwell equations found by Narozhny and Fofanov \cite{narozhny-jetp00} (referred below as NF beam).
This solution describes a family of stationary (monochromatic) focused beams which are superpositions of plane waves with wave vectors lying inside a cone with an aperture angle $2\Delta$.
The wave vectors $\veck$ are uniformly distributed in a polar angle with respect to the beam axis $z$ and can have a nontrivial distribution in angle $\alpha$ between the projection of  $\veck$ on the plane perpendicular to the $z$ axis and the $x$ axis, see Fig.1 for illustration.
This solution, given by Eqs.(5)--(14) of~\cite{narozhny-jetp00}, can describe different polarizations defined in the limit $\Delta\to 0$ corresponding to the plane wave.
For our purposes, we use a particular solution corresponding to a beam with asymptotically linear polarization along the $x$ axis and with the transversal electric field so that $E_z=0$ everywhere.
This solution can be presented in the form
$$
\vecE(\vecr,t)=\frac{iE_0\Delta^2}{2\pi (1-\cos\Delta)}\exp\bigg\{-i\omega (t-z/c))\bigg\}\times
\nonumber
$$
\beq
\times\int\limits_{-\pi}^{+\pi}d\alpha\bigg\lbrack\sin^2\alpha\vece_x-\sin\alpha\cos\alpha\vece_y\bigg\rbrack G(\nu,\chi,\Delta)~
\label{A-vec}
\eeq
where 
\beq
\nu=\frac{2\pi\Delta}{\lambda}(x\cos\alpha+y\sin\alpha)~,~~~\chi=\frac{2\pi\Delta^2}{\lambda}z
\label{nu-chi}
\eeq
and the function $G$ is determined by the integral 
\beq
G(\nu,\chi,\Delta)=2\int\limits_0^1 du f_1(u,\Delta)\exp\bigg\{i\nu f_1(u,\Delta)-i\chi f_2(u,\Delta)\bigg\}~,
\label{G}
\eeq
with
\beq
f_1(u,\Delta)=\frac{\sin\Delta u}{\Delta}~,~~~f_2(u,\Delta)=\frac{2\sin^2\bigg(\Delta u/2\bigg)}{\Delta^2}~.
\label{f1f2}
\eeq
This solution describes a field with its maximum amplitude $E_0$ at the origin, $\nu=\chi=0$, and intensity distribution shown in Fig.~2.
In order to provide a meaningful comparison of the ion yields produced in the two beams, we choose their parameters such that the maximal intensity ${\cal I}_{\rm m}$ and focal volume $V(\xi)$ calculated for the parabolic expansion near the focus center are correspondingly equal.
These requirements are met if the parameters in (\ref{I-gauss}) and (\ref{A-vec}) are connected by Eq.(\ref{I}) and
\beq
\Delta=2^{3/4}\theta~.
\label{2beams}
\eeq
Although the value of $cE_0^2/8\pi$ in (\ref{I}) is not the true intensity, it differs from that defined as the absolute value of the Poynting vector only by terms $\sim\Delta\approx 0.1$.
Besides, both solutions (\ref{I-gauss}) and (\ref{A-vec}) describe monochromatic static beams, while we devise a theory applicable for description of ionization dynamics in femtosecond laser pulses.
A time-dependent envelope can be added as a factor.
After that the solution (\ref{A-vec}) is no longer exact.
It remains approximately valid under the condition $T\gg w_0/c$ which is reasonably satisfied for the chosen $w_0=3\mu{\rm m}$ and $T=33$fs.
\begin{figure}
\includegraphics[width=9cm]{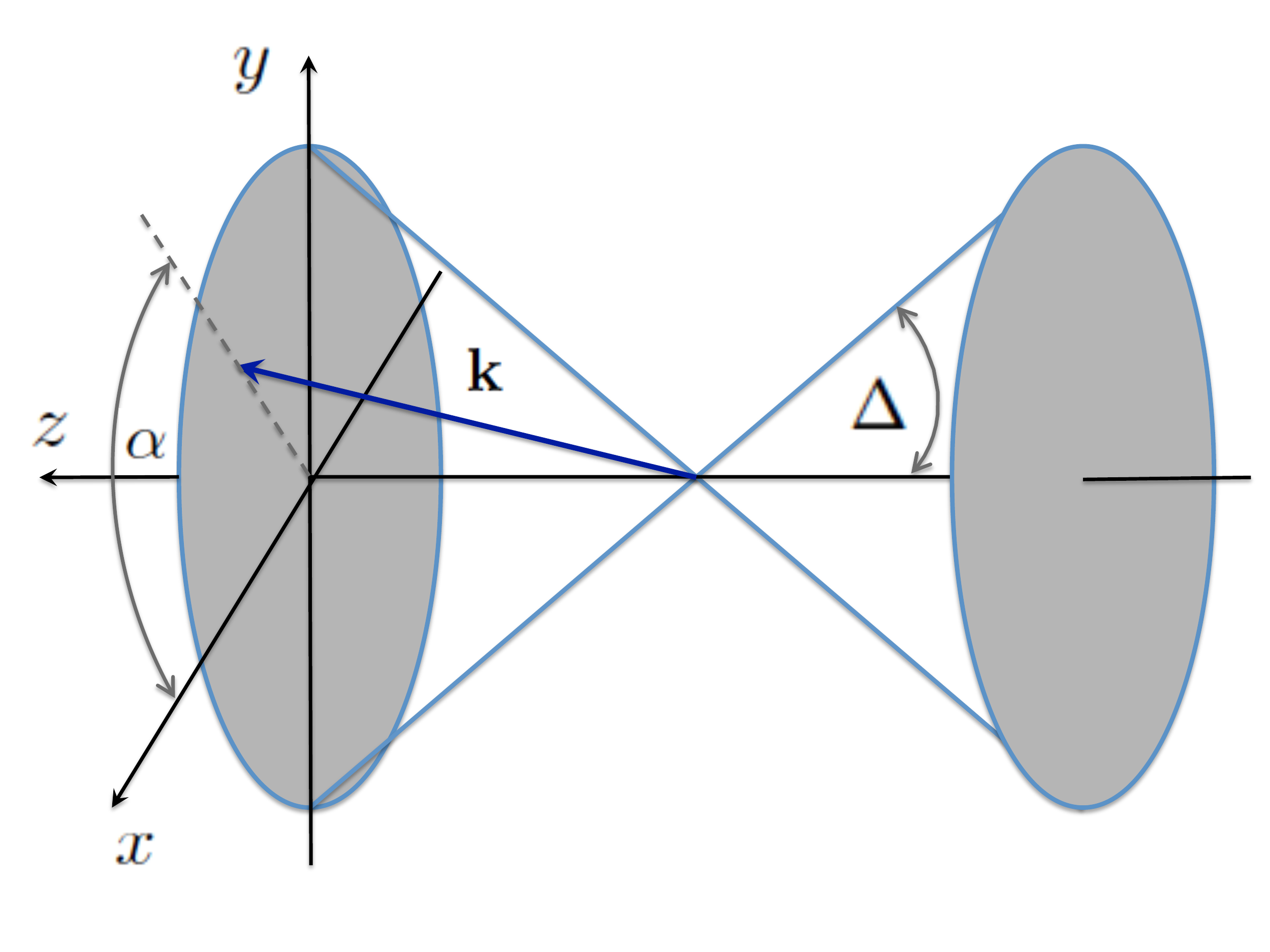}
\caption{A scheme explaining notations in Eqs.(\ref{A-vec})--(\ref{f1f2}).}  
\label{fig1} 
\end{figure}
\begin{figure}
\includegraphics[width=9cm]{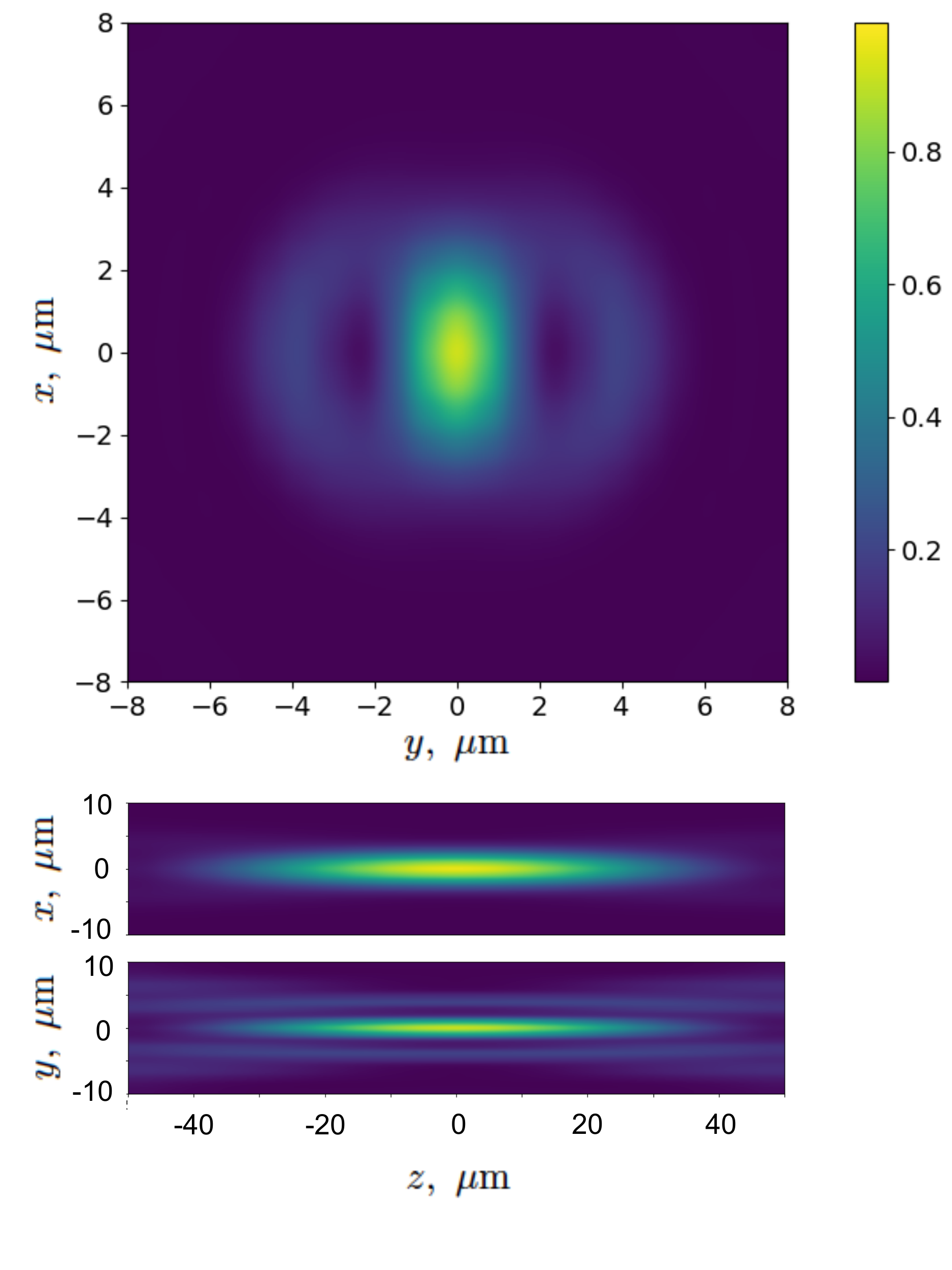}
\caption{(Color online) Normalized distributions in effective intensity (\ref{I}) in the planes $(x,y,z=0)$ (upper panel), $(x,y=0,z)$ (middle panel) and $(x=0,y,z)$ (lower panel) for the NF beam with $\Delta=0.178$ corresponding to parameters $\lambda=1\mu{\rm m}$ and $w_0=3\mu{\rm m}$ of the Gaussian beam (\ref{I-gauss}).}  
\label{fig2} 
\end{figure}

\section{Results and discussion}

Using the intensity-dependent populations $C_k$ numerically found in \cite{ciappina-pra19} we calculate the number of highly charged ions of argon, krypton and xenon in the intensity interval $3\cdot 10^{20}\div 5\cdot 10^{24}$W/cm$^2$.
Results are shown in Fig.3.
The initial neutral gas density is taken $n_0=2\cdot 10^{12}$cm$^{-3}$ corresponding to a pressure in the gas jet $p\approx 5.6\cdot 10^{-5}$Torr at room temperature.
These parameters of the gas target are numerically close to those used in the experiment of \cite{link-rsi06} where the method of atomic diagnostics was verified.
In the same experiment a TOF detector with a $\simeq 50\%$ efficiency was used to identify the highest ionic states.
It can be reasonably assumed for a high-quality detector that several dozen of ions produced in the central part of the focus would by sufficient for a reliable measurement in the single-shot regime.
Given that we set the detection threshold as $N(A^{k+})\simeq 10$.
Above this threshold, the number of ions grows by approximately one order of magnitude in a relatively narrow interval of intensity.
We denote the corresponding intensities as ${\cal I}_{10}$ and ${\cal I}_{100}$.
Their numerical values are shown in Table 1 for the two pulse shapes.
These data give for the intervals $\Delta{\cal I}={\cal I}_{100}-{\cal I}_{10}$ of intensity corresponding to a one order in magnitude growth in the number of ions $\Delta{\cal I}({\rm Ar}^{18+})\approx 0.7\cdot 10^{21}$W/cm$^2$, $\Delta{\cal I}({\rm Kr}^{34+})\approx 0.3\cdot 10^{21}$W/cm$^2$, $\Delta{\cal I}({\rm Xe}^{52+})\approx 0.2\cdot 10^{22}$W/cm$^2$ and $\Delta{\cal I}({\rm Xe}^{54+})\approx 0.5\cdot 10^{24}$W/cm$^2$.
These numbers along with those of ${\cal I}_{10}$ and ${\cal I}_{100}$ can be used to determine the peak intensity value with accuracy $\le 30\%$ by detecting the highest charge states in discernible amounts.
It is instructive to compare the values of of ${\cal I}_{10}$ and ${\cal I}_{100}$  with that introduced in \cite{ciappina-pra19} on the basis of a simple estimate made with logarithmic accuracy (see Eq.(14) there).
This equation can be reformulated as 
\beq
{\cal I}_*\approx 7.03\cdot 10^{-6}(I_{\rm p})^{3}[10^{20}]{\rm W/cm}^2~.
\label{I*}
\eeq
The values of ${\cal I}_*$ are given in Table 1 and shown in Fig.3 by vertical red dashed lines.
For ${\rm Ar}^{18+}$ the estimate of \cite{ciappina-pra19} is in a good agreement with ${\cal I}_{100}$, while for the ions of Kr and Xe a roughly $\sim 50\%$ difference is apparent.
This discrepancy is however not crucial as the estimate (\ref{I*}) is only designed to approximately define the domain of intensities which can be probed with the given charge state.

It is clearly seen that the ion yields calculated for the two pulse shapes practically coincide in the domain $N(A^{k+})=10\div 100$ where they can already be reliably detected by a sensitive TOF setup.
This makes the approach based on the identification of intensity interval corresponding to a quick grown in the ion signal within $N\sim 10\div 100$  practically insensitive to the pulse shape. 
Note that in this interval of intensities the slope of the ion yields, found in the logarithmic scale, appears considerably larger than that at higher intensities.
This feature can also be used for identification of the respective intensity intervals.
A serious drawback of this algorithm stems from the fact that the efficiency of the TOF detector and the atomic concentration $n_0$ in the gas jet might not be precisely known.
These two uncertainties will lead to that in the coefficient connecting the number of ions shown by the curves of Fig.3 and the number of counts in the TOF detector.

A complementary approach can be based on the determination of intensities for which the numbers of two charge states appear equal \cite{ciappina-lpl20}.
These points are indicated on Fig.3 by open squares (for the Gaussian beam) and open circles (for the NF beam) for the charge states ${\rm Ar}^{17+}$ and ${\rm Ar}^{18+}$, ${\rm Kr}^{27+}$ and ${\rm Kr}^{34+}$, ${\rm Xe}^{51+}$ and ${\rm Xe}^{52+}$ and ${\rm Xe}^{53+}$ and ${\rm Xe}^{54+}$ correspondingly.
Numerical values of these intensities denoted as ${\cal I}_0$ are given in Table 1.
Positions of the intersection points do not depend on $n_0$ and on the detector efficiency provided the latter is close for two ions of the chosen charge states.
However, these positions depend on the focal intensity distribution, as seen for the data of Fig.3 and Table 1.
The bigger the difference in the ionization potentials, the further the intersection point from the initial steep part of the yield curve and the larger the difference in their positions found for the two beams.

\begin{figure}
\includegraphics[width=9cm]{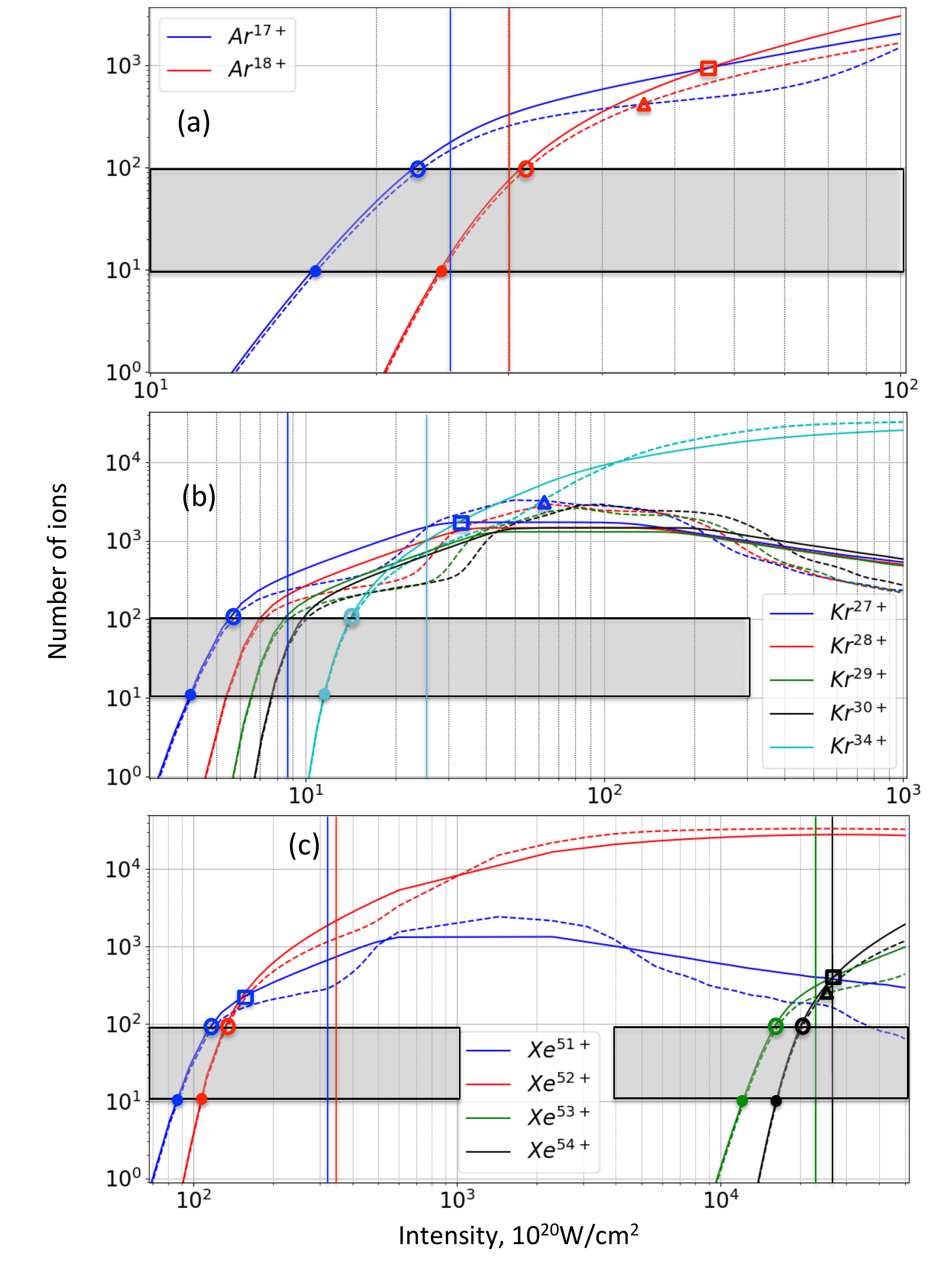}
\caption{Number of ions versus the peak laser intensity calculated along Eq.(\ref{N}) for argon (a), krypton (b) and xenon (c) using the intensity distributions for the Gaussian (solid lines) and NF (dashed lines) beams. 
The beam parameters are the same as in Fig.2.
The grey areas indicate the interval $N(A^{k+})=10\div 100$ where the number of ions grows by one order in magnitude.
The values of ${\cal I}_{10}$ and ${\cal I}_{100}$ given in Table 1 below are indicated by filled and open circles correspondingly.
The values of ${\cal I}_0$ where the numbers of two selected charge state become equal are shown by open squares for the Gaussian and by open triangles for the Narozhny-Fofanov beam.
For the pair of states ${\rm Xe}^{51+}$ -- ${\rm Xe}^{52+}$ the values of ${\cal I}_0$ for the two beams are visually close, so that the corresponding triangle sign is omitted.}
\label{fig3} 
\end{figure}

\begin{table*}[h]
\caption{\label{table:table1} Reference intensity values for ion yields shown in Fig.3: ${\cal I}_{10}$ and ${\cal I}_{100}$ -- intensities corresponding to $N(A^{k+})=10$ and $N(A^{k+})=100$ respectively; ${\cal I}_0$ is a value of intensity such that $N(A^{k+})=N(A^{(k+1)+})$ and ${\cal I}^*$ is given by the analytic estimate (\ref{I*}). Intensities are given in units $10^{20}$W/cm$^2$.
Parameters of the laser beams: $w_0=3\mu{\rm m}$, $\lambda=1\mu{\rm m}$ and $\Delta=0.178$.}

\begin{tabular}{lllllllll}

Ion & $I_{\rm p}$, eV & ${\cal I}_{10}^{\rm G}$ &  ${\cal I}_{10}^{\rm NF}$  & ${\cal I}_{100}^{\rm G}$ & ${\cal I}_{100}^{\rm NF}$ & ${\cal I}_0^{\rm G}$ & ${\cal I}_0^{\rm NF}$ & ${\cal I}^*$\\
\\
${\rm Ar}^{17+}$ & 4120 & 17 & 17 & 23 & 23 & 56 & 45 & 24\\
\\
${\rm Ar}^{18+}$ & 4426 & 24 & 25 & 31 & 32 & & & 30\\
\\
${\rm Kr}^{27+}$ & 2929 & 4.1 & 4.1 & 5.5 & 5.7 & 32 & 62 & 8.8\\
\\
${\rm Kr}^{34+}$ & 4108 & 11 & 11 & 14 & 14 &  &  & 24\\
\\
${\rm Xe}^{51+}$ & 9607 & 86 & 87 & 118 & 120 & 141 & 145 & 310\\
\\
${\rm Xe}^{52+}$ & 9812 & 110 & 110 & 130 & 130 & & & 330\\
\\
${\rm Xe}^{53+}$ & 40272 & $1.2\cdot 10^4$ & $1.2\cdot 10^4$ & $1.6\cdot 10^4$ & $1.7\cdot 10^4$ & $2.7\cdot 10^4$ & $2.5\cdot 10^4$ & $2.3\cdot 10^4$\\
\\
${\rm Xe}^{54+}$ & 41300 & $1.6\cdot 10^4$ & $1.6\cdot 10^4$ & $2.1\cdot 10^4$ & $2.0\cdot 10^4$ &  &  & $2.5\cdot 10^4$\\

\end{tabular}

\end{table*}

\section{Conclusions}

In conclusion, we have studied the effect of the focal intensity distribution on the scheme for the peak intensity determination based on the observation of multiple tunneling ionization of heavy atoms.
Two particular algorithms for extraction of the peak intensity value were examined.
The first, based on the detection of newly appeared charge states and of a steep growth in the number of ions of these states with intensity increasing was shown essentially shape-independent but suffering some uncertainty due to that in the TOF detection efficiency and atomic concentration in the gas jet.
The second algorithm requires measuring the ratio of the ion numbers for two selected charge states.
This is robust with respect to the TOF efficiency and concentration but appears generally dependent on the intensity distribution in the focal spot.
This dependence is less pronounced for pairs of charge sates with relatively close ionization potentials, e.g. for ${\rm Xe}^{51+}$ and ${\rm Xe}^{52+}$ and ${\rm Xe}^{53+}$ and ${\rm Xe}^{54+}$ correspondingly.
In an experiment targeted to measure the peak intensity value through the detection of highly charged ions produced in the process of tunneling ionization, the two algorithms can be simultaneously applied to enhance the accuracy.

We also presented a brief comparative overview of several approaches for an {\em in situ} measurement of extreme laser intensities.
We hope that our analysis will help the reader to orient in this currently emerging field of laser diagnostics.
The present paper concludes a set of publications \cite{ciappina-pra19,ciappina-lpl20} dedicated to the theoretical analysis of the atomic diagnostics method for extreme laser intensities.

\section{Acknowlegement}
Authors are grateful to S. V. Bulanov, A. Fedotov, G. Korn, E. Khazanov, A. Soloviev and S. Weber for stimulating discussions and useful suggestions. 
SVP acknowledges financial support of the Russian Foundation for Basic Research via Grant No. 19-02-00643. 
MFC acknowledges the Spanish Ministry MINECO (National Plan 15 Grant: FISICATEAMO No.~FIS2016-79508-P, FPI), European Social Fund, Fundació Cellex, Generalitat de Catalunya (AGAUR Grant No. 2017 SGR 1341, CERCA/Program), ERC AdG NOQIA, EU FEDER, and the National Science Centre, Poland-Symfonia Grant No. 2016/20/W/ST4/00314. 
This work was supported by the project 'Advanced research using high intensity laser produced photons and particles'
(CZ.02.1.01/0.0/0.0/16\_019/0000789) through the European Regional Development Fund (ADONIS).


\section*{References}
\begin{thebibliography}{9}

\bibitem{ivanov-rmp09} F. Krausz and M. Ivanov, ``Attosecond physics,'' \rmp {\bf 81}, 163 (2009).

\bibitem{agostini-adv12} L.F. DiMauro and P. Agostini, ``Atomic and Molecular Ionization Dynamics in Strong Laser Fields: From Optical to X-rays,'' Adv. At. Mol. Opt. Phys. {\bf 61}, 117 (2012).

\bibitem{karnakov-ufn15} B.M. Karnakov, V.D. Mur, S.V. Popruzhenko and V.S. Popov, ``Current progress in developing the nonlinear ionization theory of atoms and ions,'' Phys. Usp. {\bf 58} (2015).

\bibitem{bulanov-rmp09} G. Mourou, T. Tajima and S.V. Bulanov, ``Optics in the relativistic regime,'' \rmp {\bf 78}, 309 (2009).

\bibitem{dipiazza-rmp12} A. Di Piazza, C. M\"{u}ller, C.Z. Hatsagortsyan and C.H. Keitel, ``Extremely high-intensity laser interactions with fundamental quantum systems,'' \rmp {\bf 84}, 1177 (2012).

\bibitem{fedotov-cp15} N.B. Narozhny and A.M. Fedotov, ``Extreme light physics,'' Cont. Phys. {\bf 56}, 249 (2015).

\bibitem{perry-ol99} M. D. Perry \etal, ``Petawatt laser pulses,'' Opt. Lett. {\bf 24}, 160 (1999).

\bibitem{bahk-ol04} S.-W. Bahk, P. Rousseau, T. A. Planchon, V. Chvykov, G. Kalintchenko, A. Maksimchuk, G. Mourou, and V. Yanovsky, ``Generation and characterization of the highest laser intensities $(10^{22}{\rm W/cm}^2)$,'' Opt. Lett. {\bf 29}, 2837 (2004).

\bibitem{yanovsky-oe08} V. Yanovsky \etal, ``Ultra-high intensity 300-TW laser at 0.1 Hz repetition rate,'' Opt. Express {\bf 16}, 2109 (2008).

\bibitem{sulf-22} Z. Guo \etal, ``Improvement of the focusing ability by double deformable mirrors for 10-PW-level Ti:sapphire chirped pulse amplification laser system,'' Opt. Exp. {\bf 26}, 26776 (2018).

\bibitem{apollon} D. Papadopoulos \etal, ``The Apollon 10 PW laser: experimental and theoretical investigation of the temporal characteristics,'' High Pow. Las. Sci. Eng. {\bf 4} E34 (2016),  doi:10.1017/hpl.2016.34; ``First Commissioning Results of the Apollon Laser on the 1 PW Beam Line,'' 2019 Conference on Lasers and Electro-Optics (CLEO), doi:10.1364/CLEO\_SI.2019.STu3E.4.

\bibitem{APRI} J.H. Sung \etal, ``4.2??PW, 20??fs Ti:sapphire laser at 0.1??Hz,'' Opt. Lett. {\bf 42}, 2058 (2017).

\bibitem{CAEP} X. Zeng \etal, ``Multi-petawatt laser facility fully based on optical parametric chirped-pulse amplification,'' Opt. Lett. {\bf 42}, 2014 (2017).

\bibitem{SULF} Z. Gan \etal, ``200 J high efficiency Ti:sapphire chirped pulse amplifier pumped by temporal dual-pulse,'' Opt. Exp., {\bf 25} 5169 (2017); W. Li \etal, ``339??J high-energy Ti:sapphire chirped-pulse amplifier for 10??PW laser facility,'' Opt. Lett. {\bf 43}, 5681 (2018).

\bibitem{ELI} J-P. Chambaret \etal, 2010 ``Extreme light infrastructure: laser architecture and major challenges,'' in ``Solid State Lasers and Amplifiers IV, and High-Power Lasers'' {\bf 7721} 77211D

\bibitem{P3-ELI} S. Weber \etal, ``P3: An installation for high-energy density plasma physics and ultra-high intensity laser–matter interaction at ELI-Beamlines,'' Mat. Radiat. Extr. {\bf 2}, 149 (2017).

\bibitem{XCELS} A.V. Bashinov, A.A. Gonoskov, A.V. Kim, G. Mourou and A.M. Sergeev, ``New horizons for extreme light physics with mega-science project XCELS,'' Eur. Phys. J.  Spec. Top. {\bf 223}, 1105 (2014).

\bibitem{ciappina-pra19} M. F. Ciappina, S. V. Popruzhenko, S. V. Bulanov, T. Ditmire, G. Korn, and S. Weber, ``Progress toward atomic diagnostics of ultrahigh laser intensities,'' \pra {\bf 99}, 043405 (2019).

\bibitem{ciappina-book19} M. F. Ciappina, S. V. Bulanov, T. Ditmire, G. Korn, and S. Weber, in {\em Progress in Ultrafast Intense Laser Science XV} ed. by D. Charalambidis and K. Yamanouchi, Springer, in press.

\bibitem{ciappina-lpl20} M.F. Ciappina and S.V. Popruzhenko, ``Diagnostics of ultra-intense laser pulses using tunneling ionization,'' Las. Phys. Lett. {\bf 17}, 025301 (2020).

\bibitem{dipiazza-ol12} O. Har-Shemesh, A. Di Piazza, ``Peak intensity measurement of relativistic lasers via nonlinear Thomson scattering,'' Opt. Lett. {\bf 37}, 1352 (2012).

\bibitem{he-oe19} C.Z. He \etal, ``Towards an in situ, full-power gauge of the focal-volume intensity of petawatt-class lasers,'' Opt. Exp. {\bf 27}, 30020 (2019).

\bibitem{marklund-arxiv19} T.G. Blackburn, E. Gerstmayr, S.P.D. Mangles, M. Marklund, ``Model-independent inference of laser intensity,'' arXiv:1911.02349 (2019).

\bibitem{mackenroth-njp19} F. Mackenroth, A.R. Holkundkar, H.P. Schlenvoigt, ``Ultra-intense laser pulse characterization using ponderomotive electron scattering,'' New J. Phys. {\bf 21}, 123028 (2019).

\bibitem{vais-njp20} O. E. Vais, A. G. R. Thomas, A. M. Maksimchuk, K. Krushelnick, V. Yu. Bychenkov, ``Characterizing extreme laser intensities by ponderomotive acceleration of protons from rarified gas,'' New J. Phys. {\bf 22}, 023003 (2020).

\bibitem{oppen-pr28} J.R. Oppenheimer, ``Three Notes on the Quantum Theory of Aperiodic Effects,'' Phys. Rev. {\bf 31}, 66 (1928).

\bibitem{walker-pra01} C.A. Chowdhury, C.P.J. Barty and B.C. Walker, ``Nonrelativistic'' ionization of the L-shell states in argon by a ``relativistic'' $10^{19}{\rm W/cm}^2$
laser field,'' \pra {\bf 63}, 042712 (2001).

\bibitem{walker-josa03} E.A. Chowdhury and B.C. Walker, ``Multielectron ionization processes in ultrastrong laser fields,'' J. Opt. Soc. Am. B {\bf 20}, 109 (2003).

\bibitem{yamakawa-pra03} K. Yamakawa, Y. Akahane, Y. Fukuda, M. Aoyama, N. Inoue and H. Ueda, ``Ionization of many-electron atoms by ultrafast laser pulses with peak intensities greater than $10^{19}{\rm W/cm}^2$,'' \pra {\bf 68}, 065403 (2003).

\bibitem{yamakawa-jmo03} K. Yamakawa, Y. Akahane, Y. Fukuda, M. Aoyama, J. Ma, N. Inoue, H. Ueda and H. Kiriyama, ``Super strong field ionization of atoms by $10^{19}{\rm W/cm}^2$, 10 Hz laser pulses,'' \jmo {\bf 50}, 2515 (2003).

\bibitem{link-rsi06} A. Link \etal, ``Development of an {\em in situ} peak intensity measurement method for ultraintense single shot laser-plasma experiments at the Sandia
Z petawatt facility,'' Rev. Sci. Inst. {\bf 77}, 10E723 (2006).


\bibitem{narozhny-jetp00} N.B. Narozhny, M.S. Fofanov, ``Scattering of relativistic electrons by a focused laser pulse,'' JETP {\bf 90}, 753 (2000).


\bibitem{ritus-jetp64} A.I. Nikishov, V.I. Ritus, ``Quantum Processes in the Field of a Plane Electromagnetic Wave and in a Constant Field. I,'' Zh. Eksp. Teor. Fiz. {\bf 46} 776 (1964) [{\em Sov. Phys. JETP} {\bf 19} 529 (Engl. transl.)]

\bibitem{dauIV}  V. B. Berestetskii, E.M. Lifshitz, L. P. Pitaevskii {\em Quantum Electrodynamics}, Chap.IV, Sec.101 (1982  Butterworth-Heinemann) ISBN 978-0-7506-3371-0.

\bibitem{sarachik-prd70} E.S. Sarachik and G. T. Schappert, ``Classical theory of the scattering of intense laser radiation by free electrons,'' Phys. Rev. D {\bf 1}, 2738 (1970).

\bibitem{vais-apb18} O. E. Vais, and V. Yu. Bychenkov,  ``Direct electron acceleration for diagnostics of a laser pulse focused by an off-axis parabolic mirror,'' Appl. Phys. B {\bf 124} 211 (2018).


\bibitem{kibble-pr66} T. W. B. Kibble, ``Mutual refraction of electrons and photons,'' Phys. Rev. {\bf 150} 1060 (1966).

\bibitem{gor-josab89} S.P. Goreslavsky, N.B. Narozhny and V.P. Yakovlev, ``Electron spectra of above-threshold ionization in a spatially inhomogeneous field,'' J. Opt. Soc. Am. B {\bf 6} 1752 (1989);  S.P. Goreslavsky and N.B. Narozhny, ``Ponderomotive scattering at relativistic laser intensities,'' J. Nonlinear Opt. Phys. Mat. {\bf 4}, 799 (1995).

\bibitem{bauer-jpb06} D.B. Milosevi\'c, G.G. Paulus, D. Bauer and W. Becker, ``Above-threshold ionization by few-cycle pulses,'' J. Phys. B: At. Mol. Opt. Phys. {\bf 39} R203 (2006).

\bibitem{poprz-jpb14} S.V. Popruzhenko, ``Keldysh theory of strong field ionization: history, applications, difficulties and perspectives,'' \jpb {\bf 47}, 204001 (2014).

\bibitem{paulus-jpb18} W. Becker, S. P. Goreslavski, D. B. Miloñevi\'c and G G Paulus, ``The plateau in above-threshold ionization: the keystone of rescattering physics,'' \jpb {\bf 48}, 162002 (2018).

\bibitem{kuebel-jpb18} M. K\"ubel, M. Arbeiter, C. Burger, N. G. Kling, T. Pischke, R. Moshammer, T. Fennel, M. F. Kling, and B. Bergues, ``Phase- and intensity-resolved measurements of above threshold ionization by few-cycle pulses,'' \jpb {\bf 51}, 134007 (2018).

\bibitem{strelkov-oe14} V. V. Strelkov, E. M\'evel, and E. Constant, ``Short pulse carrier-envelope phase absolute single-shot measurement by photoionization of gases with a guided laser beam,'' Opt. Exp. {\bf 22}, 6239 (2014).

\bibitem{pullen-pra13} M. G. Pullen, W. C. Wallace, D. E. Laban, \etal, ``Measurement of laser intensities approaching $10^{15}{\rm W/cm}^2$ with an accuracy of $1\%$,'' \pra {\bf 87}, 053411 (2013). 

\bibitem{keldysh} L.V. Keldysh, ``Ionization in the field of a strong electromagnetic wave,''  \jetp \textbf{47} , 1945 (1964) [Sov. Phys. JETP \textbf{20} , 1307 (Engl. transl.) (1965)].

\bibitem{ppt-jetp66a} A.M. Perelomov, V.S. Popov and M.V. Terentev, ``Ionization of atoms in an alternating electric field. I,'' \jetp {\bf 50}, 1393 (1966) [Sov. Phys. JETP {\bf 23}, 924 (Engl. transl.) (1966)].

\bibitem{pp-jetp67} A.M. Perelomov and V.S. Popov, ``Ionization of atoms in an alternating electric field. III,'' \jetp {\bf 52}, 514 (1967) [Sov. Phys. JETP {\bf 25}, 336 (Engl. transl.) (1967)].

\bibitem{popov-usp04} V.S. Popov, ``Tunnel and multiphoton ionization of atoms and ions in a strong laser field (Keldysh theory),'' Usp. Fiz. Nauk {\bf 147} 921 (2004) [Phys. Usp. {\bf 47}, 855 (Engl. transl.) (2004)].

\bibitem{lin-jpb05} X.M. Tong and C.D. Lin, ``Empirical formula for static field ionization rates of atoms and molecules by lasers in the barrier-suppression regime,'' \jpb {\bf 38}, 2593 (2005).

\bibitem{kostyukov-pra18} I. Yu. Kostyukov and A. A. Golovanov, ``Field ionization in short and extremely intense laser pulses,'' \pra {\bf 98}, 043407 (2018).

\bibitem{saloman-Ar} E.B. Saloman, ``Energy levels and observed spectral lines of ionized Argon, ArII through ArXVIII,'' J. Phys. Chem. Ref. Data {\bf 39}, 033101 (2010)

\bibitem{saloman-Kr} E.B. Saloman, ``Energy Levels and Observed Spectral Lines of Krypton, Kr I through Kr XXXVI,'' J. Phys. Chem. Ref. Data {\bf 36}, 215 (2007).

\bibitem{saloman-Xe} E.B. Saloman, ``Energy Levels and Observed Spectral Lines of Xenon, Xe I through Xe LIV,'' J. Phys. Chem. Ref. Data {\bf 33}, 765 (2004).











\end{thebibliography}
\end{document}